\documentclass{article}
\usepackage{spconf,amsmath,graphicx, hyperref}
\usepackage{tikz}
\usepackage{pgfplots}
\usepackage[font=small,skip=2pt]{caption}
\usepackage{soul}
\title{StemGen: A music generation model that listens}
\name{
    Julian D. Parker\qquad
    Janne Spijkervet$^*$\qquad
    Katerina Kosta$^*$\qquad
    Furkan Yesiler\qquad\\
    Boris Kuznetsov\qquad
    Ju-Chiang Wang\qquad
    Matt Avent\qquad
    Jitong Chen\qquad
    Duc Le
}
\address{SAMI, ByteDance Inc.}
\begin{document}
\ninept
\maketitle
\def\thefootnote{*}\footnotetext{Equal contribution}

\begin{abstract}
End-to-end generation of musical audio using deep learning techniques has seen an explosion of activity recently. However, most models concentrate on generating fully mixed music in response to abstract conditioning information. In this work, we present an alternative paradigm for producing music generation models that can listen and respond to musical context. We describe how such a model can be constructed using a non-autoregressive, transformer-based model architecture and present a number of novel architectural and sampling improvements. We train the described architecture on both an open-source and a proprietary dataset. We evaluate the produced models using standard quality metrics and a new approach based on music information retrieval descriptors. The resulting model reaches the audio quality of state-of-the-art text-conditioned models, as well as exhibiting strong musical coherence with its context.
\end{abstract}

\vspace{-4mm}
\section{Introduction}
\label{sec:intro}
%\vspace{-2mm}
End-to-end generation of musical audio using deep learning techniques is a fairly new discipline, the formative work having arguably started with WaveNet~\cite{wavenet}. Recent work has shown a large jump in the quality and diversity of generated music by borrowing techniques from the image and language processing fields. Some approaches operate on tokenized audio representations~\cite{encodec, dac} using techniques from the large language model (LLM) literature~\cite{musiclm, musicgen, vampnet}. Other approaches use score-matching techniques to generate audio directly~\cite{noise2music, spectrogram_diffusion}, or encoded as a continuous latent representation~\cite{mousai}.
%While some approaches operate on tokenized audio representations~\cite{encodec} using techniques from the large language model (LLM) literature~\cite{musiclm, musicgen, vampnet}, others use score-matching techniques to generate audio directly~\cite{noise2music, spectrogram_diffusion}.

%Music can generally be thought of as the sum of a number of independent but strongly related individual parts, which are combined together to produce a full piece of music. 
Music can generally be thought of as the sum of a number of independent but strongly related individual parts, conceived at different levels of elaboration and textural layout which are combined together to produce a full piece of music
\cite[p.~194]{cook1990music}.
A part may correspond to a musician playing a particular instrument, or to something more abstract like the output of a sampler or synthesizer. These parts are often colloquially referred to as \emph{stems}. Music production can be thought of as an iterative procedure that seeks to add and refine these stems to fit the aesthetic preferences of the producer or musician. In order for the resulting mixed music to sound coherent, each of these stems must be constructed to be sympathetic with the context of the existing composition, both musically and texturally. A generative model that performs this task would be a desirable tool for those making music, and one which could support existing creative workflows instead of supplanting them.

Most existing generative models for musical audio are conditioned on abstract information, varying from text descriptions~\cite{musiclm, musicgen, noise2music} to style categories~\cite{jukebox}, and cannot be used easily for this iterative composition approach. Models that can listen to the context audio directly and generate a musically appropriate response are rarer but have been proposed in the context of generating musical accompaniments for singing~\cite{singsong} and in the context of generating multi-track instrumental music~\cite{multi_source_diffusion}. However, these models place fixed semantic restrictions on both the context audio and the generated response.

In this work we propose a novel training paradigm for producing end-to-end generative models that are able to listen to a musical context and generate an appropriate response. We propose a network architecture for implementing such a model, based on a non-autoregressive language model with a transformer backbone. This architecture is extended from previous literature~\cite{vampnet} with a novel approach to combining multiple streams of audio tokens. We introduce two novel improvements to generation: \emph{multi-source classifier-free-guidance}, and \emph{causal-bias} during iterative decoding. We train the architecture on two datasets of stem-based musical audio and evaluate the resulting models using standard objective metrics, a novel approach using an amalgamation of music information retrieval (MIR) metrics, and listening tests.
\vspace{-2mm}
\section{Modelling approach}
%\vspace{-2mm}
\label{sec:approach}
Given a dataset that contains pieces of music separated into their stems, we can construct pairs of data consisting of a \emph{context-mix}, which represents the musical context, and a \emph{target-stem}, which represents a response to that context. If the piece of music contains $M$ stems, the context-mix can be constructed as a mixture of $N < M$ stems selected at random from the full set. The target-stem can then be taken to be a single stem, selected at random from those not included in the context-mix. This procedure means that a single song with $N$ stems can result in $\frac{N}{2}(2^N-2)$ training pairs. This acts as a intrinsic form of data augmentation, and likely somewhat mitigates the scarcity of high quality music data in stem separated format.

  \begin{figure}
     \centering
     \includegraphics{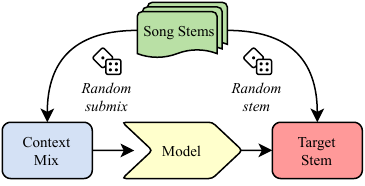}
     \caption{Schematic diagram of the StemGen training paradigm.}
     \label{fig:paradigm}
     \vspace{-4mm}
 \end{figure}
  \begin{figure*}[!h]
     \centering
     \includegraphics[scale = 1.0]{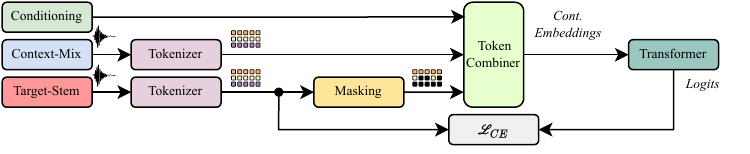}
     \caption{Schematic showing overall architecture of the StemGen model during training.}
     \label{fig:architecture}
     \vspace{-2mm}
 \end{figure*}
The training problem can be formulated in one of two ways, either as learning the joint distribution $p(\mathbf{t},\mathbf{a})$ of the context-mixes, $\mathbf{a}$, and target-stems, $\mathbf{t}$, or as learning the conditional distribution of target-stems given the context-mixes, $p(\mathbf{t} | \mathbf{a})$. We refer to modelling $p(\mathbf{t} | \mathbf{a})$ as \emph{conditional generation} and modelling $p(\mathbf{t},\mathbf{a})$ as \emph{joint generation}. In this work, we address the conditional generation problem. Fig~.\ref{fig:paradigm} shows an overview of this paradigm.
%\vspace{-4mm}
\subsection{Model Architecture}
\vspace{-2mm}
 
As with many recent models~\cite{musiclm, musicgen, vampnet}, we reformulate the problem %from modelling sampled waveforms to modelling sequences of abstract tokens.
as modelling sequences of abstract tokens rather than modelling sampled waveforms.
This allows the application of powerful techniques from the domain of language modelling. In order to produce such a sequence of tokens from audio waveforms, we use a separate audio encoding model~\cite{encodec}. This model employs a residual vector quantizer (RVQ), which means that each temporal frame of the signal is described by a set of multiple hierarchical tokens. There exist different strategies for modelling this token sequence. MusicLM~\cite{musiclm} separates the tokens into coarse and fine sets, flattens each set into a single stream with one token per-step, and then models them autoregressively. MusicGen~\cite{musicgen} also takes an autoregressive approach, but presents a number of methods for combining all the tokens for a time-step into a single sequence element. VampNet~\cite{vampnet}, which is derived from the speech model SoundStorm~\cite{soundstorm}, also combines some tokens into a single sequence element (albeit split into coarse and fine groups in a manner similar to MusicLM), but then models the resulting two sequences in a non-autoregressive manner, similar to a masked language model. Our approach is similar to MusicGen in that it employs a single model instead of separate coarse/fine models, and combines all token levels into a single sequence element. However we employ a non-autoregressive approach to training and sampling, which is more similar to VampNet. The overall structure of the architecture during training is shown in Fig.~\ref{fig:architecture}
%\subsubsection{Token and audio channel combination}

\begin{figure}[t]
     \centering
     \includegraphics{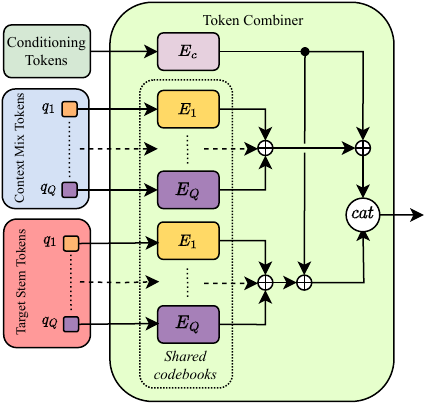}
     \caption{Schematic showing how the $Q$ RVQ levels of both the context-mix and the target stem are converted into continuous embeddings using codebooks $E_{1 \hdots Q}$, and then combined with each other along with conditioning information.}
     \label{fig:token_combiner}
 \end{figure}

 In contrast to VampNet \& MusicGen, we must combine multiple channels of audio (context-mix and target-stem) into a single sequence element. We achieve this by producing two sets of embeddings, one for each audio channel. Each embedding is produced by summing the embeddings for each hierarchical token level for that audio channel. These two embeddings are then concatenated along the embedding dimension, and used as a single sequence element.
 Before concatenation, embeddings are also calculated for any relevant non-audio conditioning. This could consist of encoded text embeddings, embeddings from a model such as CLAP~\cite{CLAP}, or embeddings produced from available categorical metadata such as instrument type or musical genre. This conditional embedding is then summed into the embeddings for each audio channel, before concatenation. Fig.~\ref{fig:token_combiner} illustrates this token combination process.

%\subsubsection{Masked training procedure}

\begin{figure}
     \centering
     \includegraphics[scale=1.2]{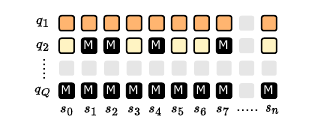}
     \caption{Example masking pattern when training to generate the 2nd level of a $Q$-level audio tokenizer. $q_{1\hdots Q}$ denote the RVQ levels, and $s_{0\hdots n}$ the sequence elements (equivalent to time steps in this case).}
     \label{fig:masking}
     \vspace{-4mm}
 \end{figure}

Training is conducted using the masking procedure of SoundStorm~\cite{soundstorm}, but with one modification \--- we ensure that when training a particular hierarchical token level, all lower hierarchical levels (corresponding to finer residuals in the RVQ) are completely masked. Fig.~\ref{fig:masking} illustrates this. This masking procedure is only applied to the tokens for the target-stem. Given that we want to train a \emph{conditional-generation} model, context-mix tokens are left completely unmasked. If we wanted to train a \emph{joint-generation} model, the context-mix tokens could be included in the masking procedure.
%Secondly, we enable training for both the \emph{joint generation} objective and the \emph{conditional generation} objective by altering the masking procedure for the source-mix channel. In the case of \emph{joint-generation}, both the source-mix and target stem are masked using the same procedure. The masked tokens for both channels are predicted using a single set of linear language-model heads (one for each hierarchical level) via a categorical cross-entropy loss. To allow the shared language-model heads to predict two different channels of tokens, the output-embeddings from the language-model backbone are passed through two seperate linear transformations, one for each audio channel. In the case of \emph{conditional generation}, the source-mix is completely un-masked, and therefore not present in the loss calculation.
%\vspace{-5mm}

\vspace{-4mm}
\subsubsection{Causally biased iterative decoding}
\begin{figure}[t]
     \centering
     \includegraphics[scale =0.9]{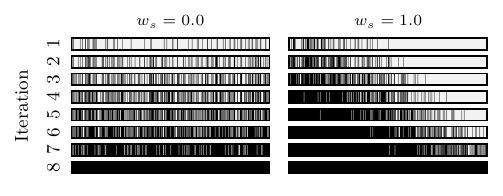}
     \caption{Example of iterative decoding for sequence of a single token level over 8 iterations, with and without causal-bias. Black denotes which sequence elements have been sampled.}
     \label{fig:causal_bias}
 \end{figure}

\vspace{-2mm}
To sample new outputs using a trained masked-LM, an iterative process is used. For a particular token level, we start with the entire sequence masked. Candidate samples are then generated for the entire sequence by sampling from the logits predicted by the backbone of the masked-LM for the current state of the sequence. These candidate samples are then ranked according to some criteria (see below), and the top-k are retained. The process is then repeated using the updated sequence, until all tokens are unmasked. It is then repeated for the next hierarchical token level. The criteria used to rank the candidate samples at each step has a strong effect on the quality of generated output. SoundStorm ranks samples by the confidence of the model, as given by the value of the logit for the sampled token. VampNet extends this by adding Gumbel noise to the confidence rankings with a defined weighting. We found that these sampling procedures produced poor output in many cases. Biasing the generation towards confidence leads to monotonous and uninteresting output, whereas relying too heavily on random selection leads to poor transients and unnatural amplitude fluttering in the output. We propose an extension to this sampling approach that encourages earlier sequence elements to be sampled first, enforcing a kind of fuzzy causality. The resulting ranking function is given by:
\begin{equation}
\rho(x_n) = w_c. c(x_n) + w_s .(1 - n/N) + w_r X
\end{equation}
where $x_n$ is the candidate sample at sequence index $n$, $N$ is the total sequence length, $c(x_n)$ is the model's confidence in the candidate sample as calculated by applying $\text{softmax}$ to the logits, $X \sim U(0,1)$ is a uniformly distributed random variable, and $w_p$, $w_s$, $w_r$ are scalar weights. In Fig.~\ref{fig:causal_bias} we show the evolution of a single sequence during iterative decoding, with and without causal-bias.
%\vspace{-3mm}
  %\begin{figure}[b] 
 \vspace{-3mm}
\subsubsection{Multi-source classifier-free-guidance}
 \vspace{-1mm}
\label{sec:multi_source}
We employ classifier-free-guidance~\cite{cfg} during the sampling of candidate tokens. In addition to applying this process to the non-audio conditioning, we also apply it to the audio context conditioning, with the hope of enforcing stronger alignment to the audio context. The classifier-free-guidance can be applied over both conditioning sources simultaneously using the standard formulation:
\begin{equation}
\log(p(\mathbf{t})p(\mathbf{c}|\mathbf{t})^\lambda) \approx \lambda \log(p(\mathbf{t} | \mathbf{c})) + (1-\lambda) \log(p(\mathbf{t}))
\end{equation}
where $\mathbf{c}$ is a combined conditioning source containing both the context-mix and any other conditioning and $\lambda$ is the guidance scale.

We also introduce a technique for weighting the guidance for the multiple conditioning sources independently:
\begin{align}
\log(p(\mathbf{t})\prod_{i=1}^N p(\mathbf{c}_i |  \mathbf{t}) ^{\lambda_i}) & \approx \sum_{i=1}^N \lambda_i \log(p(\mathbf{t} | \mathbf{c}_i) \nonumber \\ & + (1 - \sum_{i=1}^N \lambda_i) \log(p(\mathbf{t}))
\end{align}
where the $\mathbf{c}_i$ are independent conditioning sources, and $\lambda_i$ are the guidance scales for each conditioning source. The derivation follows trivially from that of single-source classifier-free guidance, via the application of Bayes' rule. Note that it is important to train the model with independent dropout for each conditioning source. 
%In the case of only two conditioning sources, this expression can be rearranged to also have a term using the full conditional logits $\log(p(\mathbf{t} | \mathbf{c}_1,\mathbf{c}_2))$. This is potentially beneficial given that this combination of conditioning is the most common during training. The full deriviation is omitted here for brevity.
\vspace{-2mm}
\section{Experiments}
\vspace{-2mm}
%\subsection{Training}
All models are trained on a single NVIDIA A100 GPU with 80GB of VRAM. The batch size is 40. Training proceeds until validation accuracy has clearly stopped increasing for 20k steps, which was generally after around 150k total training steps. This translates to around 4 days of training time. The AdamW optimizer is used, with a tri-stage learning rate schedule consisting of 10k warmup steps, 50k steps at the maximum learning rate of 0.0005, and then a slow exponential decay over 300k steps. 
%\subsubsection{Data}

As a baseline, we train on the Slakh~\cite{slakh} dataset, which consists of 145 hours of synthetic musical audio separated into stems. In addition, we train on an internal dataset of 500 hours of licensed human-played music separated into stems. Individual songs were chunked into 1 minute segments during data preparation. When constructing training pairs, a random 20s crop of each segment is chosen and the random submixing procedure described above in Sec.~\ref{sec:approach} is applied. The resulting context-mix and target-stem are checked to make sure they are not silent, and discarded if so. Both datasets contain metadata specifying the instrument type of each stem, with classes corresponding to the 18 General MIDI instrument categories~\cite{generalMIDI}. We use a trainable codebook with 18 entries to translate this conditioning information into an embedding. This model therefore has two conditioning sources: the context-mix and the instrument category, referred to as $\mathbf{c}_a$ and $\mathbf{c}_i$ respectively.
%\subsubsection{Audio tokenizer}

We use the publicly available checkpoint of Encodec~\cite{encodec} also used for MusicGen~\cite{musicgen}. This codec operates on 32kHz audio, and datasets are resampled appropriately. The encoded audio has a frame rate of 50Hz, with four tokens in the range $0\dots2047$ per frame.
%\subsubsection{Decoder backbone}

As with any LM-like structure, the above described model architecture is broadly agnostic to the network type used for the main decoder block. In practice, most models of this type employ a transformer. We follow this convention and employ a LLaMa-type transformer~\cite{llama} for the experiments presented here. The transformer has an embedding dimension of 1024 and the number of attention heads is 16. 20 layers are used, giving a total parameter count of $\sim250$M.
\vspace{-6mm}
\subsection{Evaluation Metrics}

We analyze the audio samples generated by our system in two objective evaluation strands. The first is Fréchet Audio Distance (FAD)~\cite{fad}, which we calculate on VGGish embeddings as per the original paper (referred to in results as \emph{FAD}).
%{\color{red}{It has been shown that FAD-like objective metrics in the audio domain do not necessarily correlate with comparative listening test results~\cite{vinay2022evaluating}.}} To this end, 
The second is the Music Information Retrieval Descriptor Distance (MIRDD). Similarly to FAD, MIRDD is calculated on two populations. Instead of comparing distributions in an abstract embedding space, MIRDD compares the distributions with respect to a number of MIR (pitch-,rhythm-, and structural-based) descriptors. 

Most of the descriptors are inspired by the research for melody expectation and prediction. The use of such musical quantities has been shown to enhance melodic expectation when embodied in a cognitively plausible system for music prediction and demonstrated as a successful system for evaluating objectively music generation models that fit subjective input (for more details see \cite{pearce2005construction, pearce2}). Descriptors have been developed for evaluating monophonic generations from composition models in MIDI format~\cite{lerch_eval}. We expand the concept by introducing a set designed to evaluate generated audio in both monophonic and polyphonic manner.

% Surprise and repetition in music are measurable elements that relate to expectation and memorability, both of these concepts affecting the predictability and uncertainty in the pleasure of music (for more details see~\cite{IDYOM, FANTASTIC}. 

% : From  "\cite{pearce1} demonstrate a relationship between prediction performance and fit to human expectancy data (Manzara et al., 1992), suggesting
% that human perceptual systems base their predictions on
% uncertainty-reducing representational features. In terms
% of model selection for music generation, highly predictive
% theories of a musical style, as measured by information
% content, should generate original and acceptable works in
% the style (Conklin and Witten, 1995)". ({\it{this is the evidence about comparing the statistics with the training data}}).

% Tools and techniques for automatically extracting musical information from audio signals matured as a result of advances in the MIR field.

We compute the descriptors for every audio in the reference and test populations, utilising the outcome of the models for audio transcription~\cite{MIR_transcription}, beat and downbeat detection~\cite{MIR_beat}, key and chords estimation~\cite{MIR_key_chords}, as well as structure extraction~\cite{MIR_structure}. This gives us a set of vectors of real values (for continuous quantities) or integer counts (for discrete quantities). 
%We convert the continuous features to a discrete representation by introducing data bins. 
%A histogram of each statistic is first computed per collection of generated stems and the ground-truth data, and then normalised by the count of data points; there can be a single point or multiple points per audio file, merging all counts in the latter case, effectively flattening the single-piece individuality.
MIRDD is computed, extracting the KL-Divergence of the probability distribution pairs per descriptor and averaging the values. A lower MIRDD score is considered better.
The list of descriptors used is the following: key signature, length of note pitch range, amount of unique pitch classes, maximum vertical density of percussive notes (computed in a 0.6s window size), number of beats per bar, chord labels variety, chord tonality alteration ratio (moving from minor to major and vice versa), note pitches, note pitch classes, chord triad labels and structure labels.
\vspace{-2mm}
\subsection{Results}
%\vspace{-2mm}
For evaluation, we generated sets of 400 example outputs from each model. The examples were produced following a similar procedure to that used for the construction of training examples, but pulling from a separate set of test data not contained within the training set. For each example we generate a context-mix, and randomly choose a target-stem category. The generated stem produced by the model for this set of conditioning is placed into one population. The real stem for the matching target category is placed into another population. These two populations can then be compared with a variety of metrics. This procedure ensures there is no systematic error introduced by an imbalance in target categories between the two populations. We do not perform pairwise evaluation, but instead compare the distributions of the two populations. FAD-type metrics are calculated between the two populations of isolated stems, whereas MIRDD is calculated on modified populations consisting of stems summed with their original context-mix. This allows the MIRDD metric to better penalize poor musical alignment and coherence.
\vspace{-4mm}
\subsubsection{Sampling hyper-parameters \& ablations}
\label{sec:ablation}
In order to validate our novel improvements to the decoding procedure, we perform two ablations. For these ablations we use the model trained on the Slakh dataset. Firstly, we test the impact of multi-source classifier-free-guidance, as described in Sec.~\ref{sec:multi_source}, by calculating metrics over a variety of guidance scales, $\lambda_a$ and $\lambda_i$, with respect to our two conditioning sources $\mathbf{c}_a$ and $\mathbf{c}_i$. The results can be seen in Tab.~\ref{tab:cfg}. A value of 1.0 for guidance scale is equivalent to no guidance with respect to that conditioning source. The exact best values for guidance scale are very dependent on the conditioning information, so for cases where the guidance scale is greater than 1.0 we show results averaged over 1000 examples at different values of $\lambda_i$, $\lambda_a$ up to a maximum of 4.0. The results confirm that adding guidance over multiple sources is beneficial. We settle on $\lambda_a = \lambda_i = 3.0$ as a general setting for evaluation.

\begin{table}[!h]
\begin{center}
\begin{tabular}{r|c c c c}
& FAD & MIRDD \\ \hline \hline
$\lambda_i = \lambda_a = 1.0$ & 4.30 & 0.41 \\ \hline
$\lambda_i > 1.0, \lambda_a = 1.0$ & 3.66 & 0.26 \\ \hline
$\lambda_i = 1.0, \lambda_a > 1.0$ & 3.20 & 0.29 \\ \hline
$\lambda_i > 1.0, \lambda_a > 1.0$ & \textbf{3.17}  & \textbf{0.25} 
\end{tabular}
\caption{Averaged objective metrics with different ranges of guidance scales.}
\label{tab:cfg}
\end{center}
%\vspace{-5mm}
\end{table}

Secondly, we test the impact of causal bias during iterative decoding by comparing various relative strengths of causal-bias. The other ranking weights, $w_c$ and $w_r$ are set to $0.1$ and $1.0$ respectively. A casual-bias weight $w_s$ of $0$ gives a decoding scheme similar to that used in VampNet~\cite{vampnet}. The results can be seen in Tab.~\ref{tab:causal}. We can see that adding a small amount of causal-bias has a positive effect on FAD and MIRDD metrics, indicating an increase in sound quality and musical alignment. This lessens as $w_s$ increases further. We therefore settle on $w_s = 0.1$ for further experiments.
\begin{table}[!h]
\begin{center}
\begin{tabular}{r|c c c c}
$w_s$& 0.0 & 0.1 & 0.2 & 0.5 \\ \hline \hline
FAD & 3.18 & \textbf{3.12} & 3.13 & 3.17 \\ \hline
%FAD-CLAP & 0.19 & \textbf{\color{red}{0.23}} & 0.20 & 0.21 \\ \hline
MIRDD & 0.32 & \textbf{0.14} & 0.18 & 0.20 \\ 
\end{tabular}
\caption{Objective evaluation metrics for various values of causal-bias weight $w_s$.}
\label{tab:causal}
%\vspace{-6mm}
\end{center}
\end{table}

During iterative decoding, we use 128, 64, 32 and 32 steps respectively for the four hierarchical levels of the tokenizer. This is universal across all results presented here.
\vspace{-4mm}
\subsubsection{Final evaluation}

We evaluate both the model trained on Slakh, and that trained on our internal dataset. Both models are sampled from using a \emph{best} set of sampling parameters derived above in Sec.~\ref{sec:ablation}. Additionally we sampled from each model using a set of \emph{naive} sampling parameters, which are equivalent to removing classifier-free-guidance and causal-bias in decoding. We show objective metrics for these sets of outputs in Tab.~\ref{tab:results}. Whilst a direct comparison is not possible due to the different task, the FAD scores for the models trained on both Slakh and the internal dataset are consistent with those seen for state-of-the-art text-conditioned models~\cite{musicgen}. It is also clear that the larger size and human-content of the internal dataset leads to an appreciable improvement in output quality. Examples from these models can be heard at the website associated with this work \footnote{\url{https://julian-parker.github.io/stemgen/}}
\begin{table}[!h]
\begin{center}
%\vspace{-3mm}
\begin{tabular}{r|c c c c}
& FAD  & MIRDD \\ \hline \hline
Slakh (\emph{naive}) & 4.3 & 0.43 \\ \hline
Slakh (\emph{best}) & 3.12  & 0.14 \\ \hline
Internal (\emph{naive}) & 3.39 & 0.40 \\ \hline
Internal (\emph{best}) & 1.96 & 0.36 \\ 
\end{tabular}
\caption{Objective evaluation metrics for both models with \emph{best} and \emph{naive} sampling parameters.}
\label{tab:results}
\end{center}
%\vspace{-8mm}
\end{table}

% \begin{center}
We also conducted a listening test in the Mean Opinion Score (MOS) format, by asking 10 participants with music training to verify the subjective quality of the produced model. We constructed three sets of outputs by mixing generated or real stems with their corresponding context-mix. The generated stems were taken from the \emph{naive} and \emph{best} sets of output from the model trained on the internal dataset as evaluated in Tab.~\ref{tab:results}. The real stems were taken from the references sets used for previous evaluations. We collated a set of 60 mixes using this technique (equally split between \emph{naive}, \emph{best} and real) and asked listeners to rate the overall quality on a Likert scaled from very bad (1) to very good (5). The results are shown in Table~\ref{tab:mos}, and confirm that the proposed model with \emph{best} sampling parameters is capable of creating plausible musical outcomes.
\begin{table}[!h]
\begin{center}
\begin{tabular}{r|c c c}
& Real & \emph{Best} & \emph{Naive}  \\ \hline \hline
MOS & 3.64 & 3.46 & 2.89 \\ 
\end{tabular}
\end{center}
\caption{MOS test results for real stems, and generated stems for the model trained on the internal dataset with naive and best sampling parameters.}
\label{tab:mos}
%\vspace{-10mm}
\end{table}

%%%%% MIRDD detailed values for Table 3:
% karaoke naive:
% "KL_values": {
%             "d1": 0.122,
%             "d3": 0.599,
%             "d5": 0.041,
%             "d7": 0.01,
%             "d9": 1.684,
%             "d15": 0.783,
%             "d20": 0.0,
%             "d24": 0.002,
%             "d26": 0.07,
%             "d28": 1.061,
%             "d30": 0.002
%         }

% karaoke best:
% "KL_values": {
%     "d1": 0.229,
%     "d3": 0.876,
%     "d5": 0.018,
%     "d7": 0.024999999999999998,
%     "d9": 1.292,
%     "d15": 0.45099999999999996,
%     "d20": 0.0,
%     "d24": 0.002,
%     "d26": 0.147,
%     "d28": 0.8,
%     "d30": 0.006
% }

% Slakh best:
% "KL_values": {
%     "d1": 0.198,
%     "d3": 0.266,
%     "d5": 0.020999999999999998,
%     "d7": 0.022,
%     "d9": 0.59,
%     "d15": 0.14800000000000002,
%     "d20": 0.002,
%     "d24": 0.008,
%     "d26": 0.055,
%     "d28": 0.177,
%     "d30": 0.022
% }

% Slakh naive:
% "KL_values": {
%     "d1": 1.914,
%     "d3": 0.317,
%     "d5": 0.020999999999999998,
%     "d7": 0.01,
%     "d9": 0.671,
%     "d15": 0.187,
%     "d20": 0.014,
%     "d24": 0.002,
%     "d26": 0.081,
%     "d28": 0.29600000000000004,
%     "d30": 0.019999999999999997
% }

%key signature: d1, 
%length of note pitch range: d9, 
%amount of unique pitch classes: 26, 
%maximum vertical density of percussive notes (computed in a 0.6s window size): d15,
%number of beats per bar: d20, 
%chord labels variety: d3, 
%chord tonality alteration ratio (moving from minor to major and vice versa): d28, 
%note pitches: d7, 
%note pitch classes: d24, 
%chord triad labels: d30,
%structure labels: d5

\vspace{-4mm}
\section{Conclusions}
%\vspace{-2mm}
In this work we defined a framework for training contextual music generation models using datasets of music split into stems. We presented a non-autoregressive language model\--like architecture for implementing this, along with novel changes to support multiple audio channels, and novel sampling methods. We trained this architecture on two datasets, and presented ablations that verify the impact of our novel sampling changes. We evaluate the final models using objective metrics, a novel MIR-based metric, and with a subjective listening test. This evaluation shows that the audio quality of the resulting model is competitive with state-of-the-art music generation models, and that the musical alignment with context is good.

%\vfill\pagebreak

% \section{REFERENCES}
% \label{sec:refs}
% literature.

%~\cite{*}
\bibliographystyle{IEEEbib}
%\vspace{-5mm}
\small
\bibliography{refs}

\end{document}